\documentclass[preprint,12pt]{aastex}

\newcommand  \kms   {km~s$^{-1}$} 

\newcommand  \sii   {[S~{\sc ii}]}

\newcommand  \ha    {H$\alpha$} 
\newcommand  \hb    {H$\beta$}
\newcommand  \hii   {H\,{\sc ii}} 

\newcommand  \nii   {[N\,{\sc ii}]}

\begin{document}

\title{A Critical Examination of Hypernova Remnant Candidates\\
in M101. II. NGC\,5471B}
\author{C.-H. Rosie Chen\altaffilmark{1}, You-Hua Chu\altaffilmark{1}, 
Robert Gruendl\altaffilmark{1}, and Shih-Ping Lai\altaffilmark{2}\\}
\affil{Department of Astronomy, University of Illinois, 1002 W.
Green St., Urbana, IL~61801; c-chen@astro.uiuc.edu, chu@astro.uiuc.edu,
gruendl@astro.uiuc.edu, slai@astro.uiuc.edu}
\and
\author{Q. Daniel Wang}
\affil{Department of Astronomy, University of Massachusetts, 
B-524 LGRT, Amherst, MA~01003; wqd@astro.umass.edu}
\altaffiltext{1}{Visiting astronomer, Kitt Peak National Observatory, 
National Optical Astronomy Observatories, operated by 
the Association of Universities for Research in Astronomy, Inc., 
under a cooperative agreement with the National Science Foundation.}
\altaffiltext{2}{Current address: Jet Propulsion Lab, MS 169-506,
4800 Oak Grove Dr., Pasadena, CA 91109; slai@thisvi.jpl.nasa.gov}

\begin{abstract}

NGC\,5471B has been suggested to contain a hypernova remnant because 
of its extraordinarily bright X-ray emission.  To assess its true nature, 
we have obtained high-resolution images in continuum bands and nebular 
lines with the {\it Hubble Space Telescope}, and high-dispersion long-slit 
spectra with the Kitt Peak National Observatory 4-m echelle spectrograph.
The images reveal three supernova remnant (SNR) candidates in the
giant \hii\ region NGC\,5471, with the brightest one being the 
$77 \times 60$ pc shell in NGC\,5471B.  
The \ha\ velocity profile of NGC\,5471B can be decomposed into a narrow 
component (FWHM = 41 \kms) from the background \hii\ region and a broad 
component (FWHM = 148 \kms) from the SNR shell.  
Using the brightness ratio of the broad to narrow components and the \ha\
flux measured from the WFPC2 \ha\ image, we derive an \ha\ luminosity of
$(1.4 \pm 0.1) \times 10^{39}$~ergs~s$^{-1}$ for the SNR shell.  The \sii\ 
$\lambda$6716/$\lambda$6731 doublet ratio of the broad velocity component 
is used to derive an electron density of $\sim700$ cm$^{-3}$ in the SNR shell.  
The mass of the SNR shell is thus $4600\pm500$ M$_\odot$.  With a 
$\sim330$ \kms\ expansion velocity implied by the extreme velocity extent 
of the broad component, the kinetic energy of the SNR shell is determined
to be $5 \times 10^{51}$ ergs.  This requires an explosion energy 
greater than 10$^{52}$ ergs, which can be provided by one hypernova or 
multiple supernovae.  Comparing to SNRs in nearby active star formation 
regions, the SNR shell in NGC\,5471B appears truly unique and energetic.
We conclude that the optical observations support the 
existence of a hypernova remnant in NGC\,5471B.

\end{abstract}


\newpage

\section{Introduction}

Two supernova remnants (SNRs), MF\,83 \citep{MF97} and 
NGC\,5471B \citep{Sk85}, in the giant spiral galaxy M101
have been recently suggested to be candidates for 
``hypernova remnants" \citep{Wa99}, because their explosion 
energies appear to be 1--2 orders of magnitude higher than 
the canonical 10$^{51}$ ergs explosion energy for normal 
SNRs \citep{Jo98}.  Such high explosion energies are rivaled 
only by the ``hypernovae" proposed by \citet{Pa98} and 
\citet{FW98} to explain the gamma-ray bursts (GRBs).  
As the explosion energies of MF\,83 and 
NGC\,5471B were derived from their X-ray luminosities 
measured from $ROSAT$ observations which did not resolve 
these sources spatially, it is of interest to investigate 
the nature of these two hypernova remnant candidates at 
optical wavelengths with high angular and spectral 
resolutions.

The analysis of optical and X-ray observations of MF\,83 has 
been reported by \citet[][hereafter Paper~I]{La01}.
It is shown that MF\,83 is a superbubble around an OB association 
in which multiple supernova explosions are likely to have taken place. 
Optical observations do not provide compelling evidence for a single, 
super-energetic supernova explosion.  Furthermore, {\it Chandra} 
X-ray observation of M101 shows that MF\,83 is dominated by a 
variable point source \citep{Sn01}.  
Therefore, it is unlikely that MF\,83 is a hypernova remnant.

The hypernova remnant candidate NGC\,5471B was initially identified 
and confirmed as an SNR in the B-component of the giant \hii\
region NGC\,5471 (hence the name NGC\,5471B) by \citet{Sk85} 
based on its nonthermal radio emission and high \sii/\ha\ ratio.
The shocked, high-velocity gas in NGC\,5471B was detected by 
\citet{CK86}, who concluded that the SNR 
was supermassive.  The X-ray emission from the giant \hii\ 
region NGC\,5471 was reported by \citet{WC95}, 
but the association of X-ray emission with the SNR per se was  
made by Wang (1999).

NGC\,5471B is an unusual SNR in an active star formation region. 
To study the giant \hii\ region NGC\,5471 and the SNR NGC\,5471B,
we have obtained {\em Hubble Space Telescope (HST)} Wide Field and 
Planetary Camera 2 (WFPC2) images of NGC\,5471 in emission lines
and continuum bands.  These images are used to search for SNR 
candidates in NGC\,5471, and to examine the physical structure and 
stellar/interstellar environment of the SNR NGC\,5471B.  We have 
also obtained new echelle spectra of NGC\,5471B with the 4~m telescope
at Kitt Peak National Observatory (KPNO) in order to kinematically 
separate the SNR from the background \hii\ region.  This paper 
reports these observations (\S 2), the SNR candidates identified in 
NGC\,5471 (\S3), a detailed analysis of the physical properties
of NGC\,5471B (\S4), and a discussion of the nature of NGC\,5471B
as a hypernova remnant (\S5).

\section{Observations \& Data Reduction}

\subsection{{\it HST} WFPC2 Images}

The {\it HST} WFPC2 images of NGC\,5471 were obtained on 1997 
November 1 for the Cycle 6 program GO-6928.  The observations 
were made through two emission-line filters, {\it F656N} (\ha) and 
{\it F673N} (\sii), and two continuum filters, {\it F547M}
(Str\"omgren $y$) and {\it F675W} (WFPC2 $R$) \citep{Bi96}.
The journal of observations is given in Table~\ref{tbl:log}.

The calibrated WFPC2 images, produced by the standard {\it HST}
pipeline processes, were further reduced with the IRAF and STSDAS 
routines.  The images were all corrected for the intensity- and 
position-dependent charge transfer efficiency (CTE) by applying 
a linear ramp with a correction factor chosen according to the 
average counts of the sky background \citep{Ho95}.
The images taken with the same filter were then
combined to remove cosmic rays and to produce a total-exposure map.  

To extract the \ha\ and \sii\ line fluxes from the emission line 
images, it is necessary to subtract the continuum emission.  As 
the giant \hii\ region NGC\,5471 is very luminous, its \ha\ emission 
contributes significantly to the broad R-band ({\it F675W}) image.  
Therefore, we first subtracted a scaled \ha\ image from the {\it F675W} 
image, then scaled the resultant image and subtracted it from the 
line images.  The scaling took into account the exposure times
and the filter transmission curves \citep{Bi96}. 

The continuum-subtracted \ha\ and \sii\ images were 
processed with the standard procedures \citep{Ch00}
to extract line fluxes and to produce the \sii/\ha\ ratio map.  
Due to the $\sim300$ \kms\ redshift of M101, two corrections 
have been applied to the extracted \ha\ flux.  First, as the 
filter transmission of the red-shifted \ha\ line is $\sim93$\% 
of the peak transmission, the extracted \ha\ flux is multiplied by
a correction factor of 1.07.  Second, the \nii\ $\lambda$6548 line
is red-shifted into the \ha\ bandpass at $\sim91$\% of the peak 
transmission.  Adopting the de-reddened \nii\ $\lambda$6548+6584/\hb\ 
ratio of 0.23 and the logarithmic extinction at \hb, $c$(\hb), of 
0.23 for NGC\,5471B reported by \citet{KG96}, we find that the 
\nii\ $\lambda$6548 line contributes
to $\sim2$\% of the total transmitted flux; therefore, an
additional correction factor of 0.98 is applied to the extracted 
\ha\ flux.

\begin{figure}
\epsscale{0.7}
\plotone{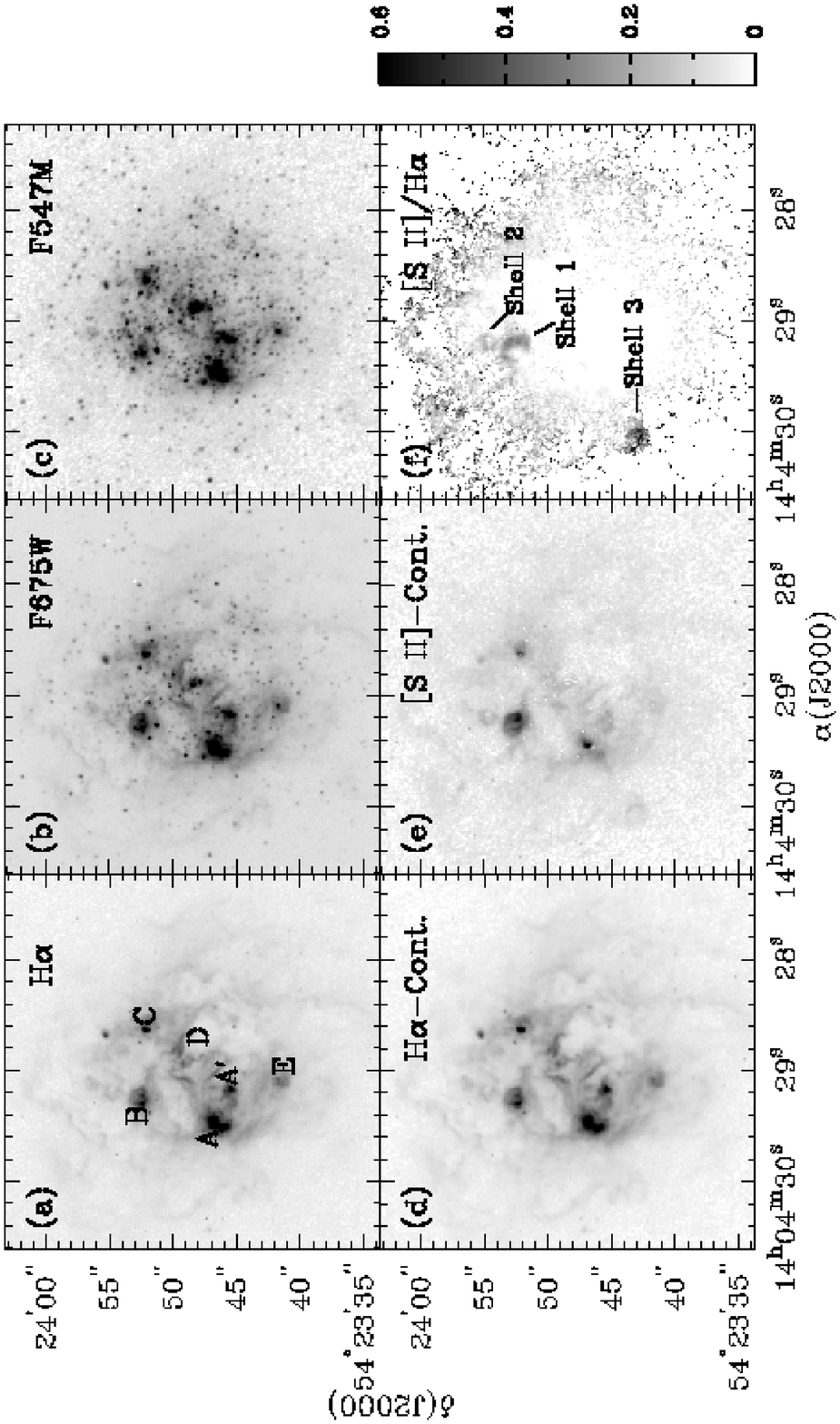}
\caption{$HST$ WFPC2 images of NGC\,5471 in (a) \ha, (b) 
{\it F675W}, (c) {\it F547M},  (d) continuum-subtracted \ha, 
(e) continuum-subtracted \sii, and (f) \sii/\ha.  
The grey-scale images in (a)--(e) are 
displayed with square-root image transfer functions.  
The \sii/\ha\ ratio map in (f) is displayed with a linear image 
transfer function, as shown in the greyscale bar to its right.}
\label{fig:ngc5471}
\end{figure}

\begin{figure}
\plotone{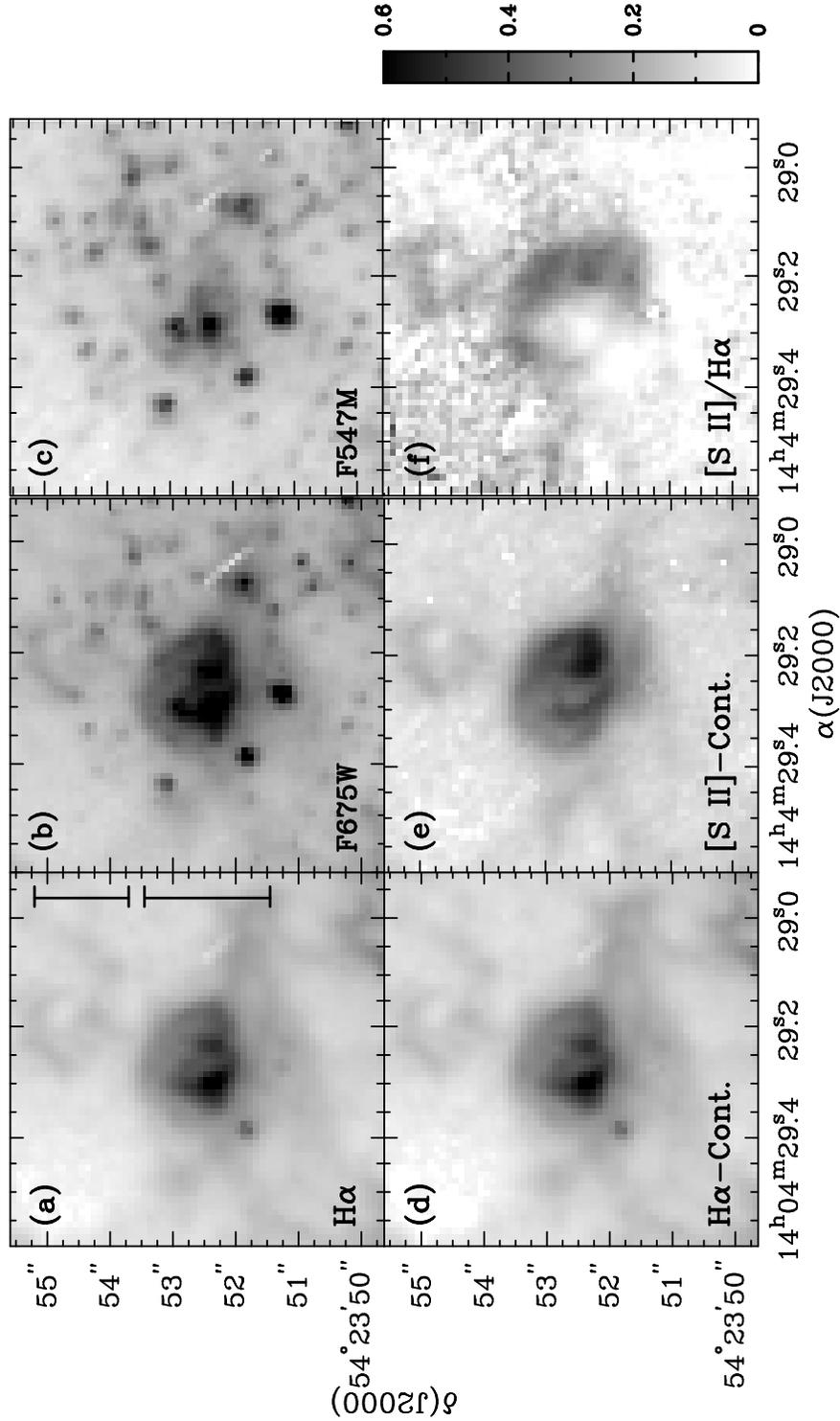}
\caption{$HST$ WFPC2 images of NGC\,5471B in (a) \ha, (b) 
{\it F675W}, (c) {\it F547M},  (d) continuum-subtracted \ha, 
(e) continuum-subtracted \sii, and (f) \sii/\ha. 
The grey-scale images in (a)--(e) are 
displayed with square-root image transfer functions.  
The \sii/\ha\ ratio map in (f) is displayed with a linear image 
transfer function, as shown in the greyscale bar to its right.
The lines in (a) mark the widths and locations of the E-W
oriented slit of the echelle observations.}
\label{fig:ngc5471b}
\end{figure}

To suppress artificial fluctuations in the \sii/\ha\ ratio in regions 
where the nebular emission is faint (and hence noisy), we have 
clipped pixels with fluxes less than 3$\sigma$ above the sky 
background for both the \ha\ and \sii\ images of NGC\,5471. 
The \ha, {\it F675W}, {\it F547M}, continuum-subtracted \ha,
continuum-subtracted \sii, and \sii/\ha\ images of NGC\,5471
are presented in Figure~\ref{fig:ngc5471}.  The A--E components
of NGC\,5471 defined by \citet{Sk85} are marked in 
Fig.~\ref{fig:ngc5471}a, and the \sii-bright shells are
marked in Fig.~\ref{fig:ngc5471}f.  Corresponding close-up 
images of NGC\,5471B are shown in Figure~\ref{fig:ngc5471b}.

\begin{figure}
\epsscale{1.0}
\plotone{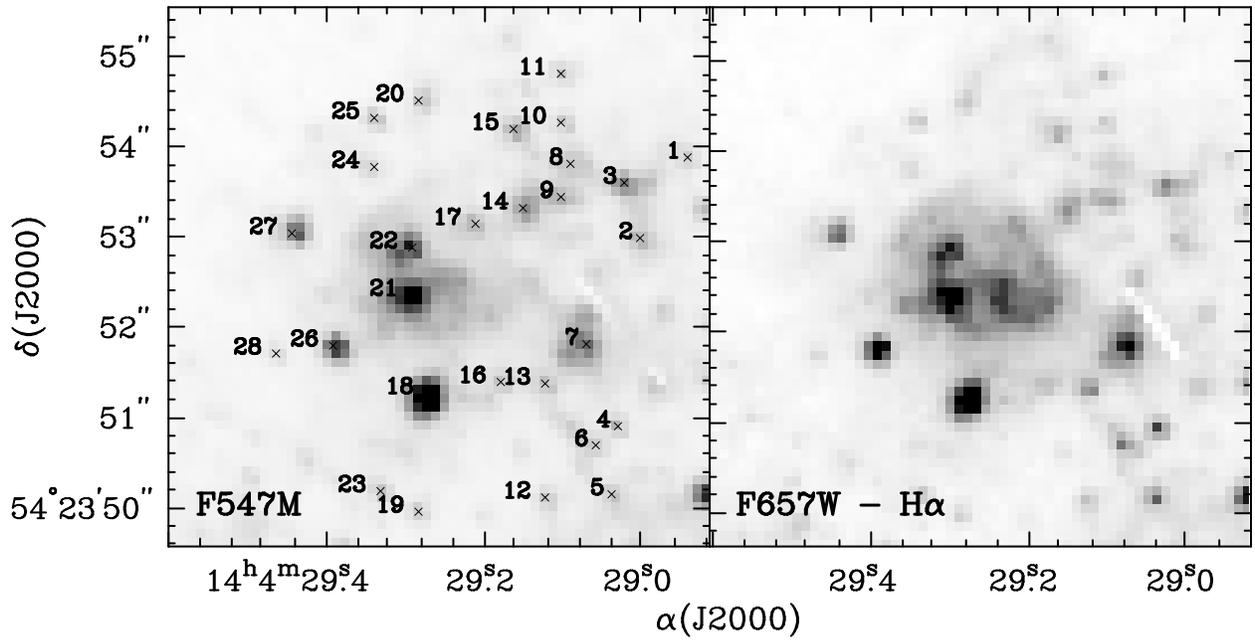}
\caption{$HST$ WFPC2 images of NGC\,5471B in (a) {\it F547M} and (b)
\ha-subtracted {\it F657W}.  Stellar sources are identified and marked
in (a)}
\label{fig:stars}
\end{figure}

Stellar photometry was carried out using the APPHOT package in IRAF
for the {\it F547M} and {\it F675W} images of NGC\,5471B.  Because of the 
complexity in the stellar field, we manually selected stellar sources
in NGC\,5471B and its surroundings and derived their apparent
magnitudes, $m_{\it F547M}$ and $m_{\it F675W}$.
The uncertainties in the photometric measurements formally given 
by APPHOT are 0.1--0.2 mag.  In most cases the true errors must 
be larger, if the uncertainties in aperture correction and sky 
background are included.  The photometric uncertainties are generally 
larger in $m_{\it F675W}$ than in $m_{\it F547M}$, because the nebular emission 
produces rapid variations in the background in the {\it F675W} image.  
Varying the background aperture may result in changes in $m_{\it F675W}$ 
by $\sim0.2$ mag for most stars, and up to $\sim0.5$ mag for 
stars near bright nebulosities.
To reduce the errors introduced by the nebular contamination in
the {\it F675W} band, we have also carried out stellar photometry for the 
\ha-subtracted {\it F675W} image, and designated this apparent magnitude as 
$m_{\it F675W'}$.  
Table~\ref{tbl:photos} lists the resulting magnitudes and 
colors of the stellar sources in NGC\,5471B.
The stellar sources are marked on the {\it F547M} image in
Figure~\ref{fig:stars}a, and the \ha-subtracted {\it F657W} image
is shown in Fig.~\ref{fig:stars}b.

\subsection{KPNO 4~m Echelle Observations}

High-dispersion spectra of NGC\,5471 were obtained with the echelle 
spectrograph on the KPNO 4~m telescope on 1999 June 30 and July 1.  
The observations were made with the 79-63 echelle grating 
and the 226-1 cross disperser. The $2048 \times 2048$ T2KB CCD 
detector was used to record the spectra.  The pixel size is 
24 $\mu$m.  To reduce the readout noise, the data were read out 
with the pixels binned by a factor of 2 in the spatial direction.
The final data array thus has a scale of 0\farcs57 pixel$^{-1}$ 
along the slit, and $\sim0.08$\AA\ pixel$^{-1}$ along the dispersion.

Two slit positions were observed, one centered on the hypernova 
remnant candidate NGC\,5471B (= Shell~1 in Fig.~\ref{fig:ngc5471}f) 
and the other on a \sii-bright shell to its north (Shell~2 in 
Fig.~\ref{fig:ngc5471}f).  For both observations, the slit was 
E-W oriented with a slitwidth large enough to cover the entire 
shell, 2\farcs0 for NGC\,5471B and 1\farcs5 for Shell~2.  NGC\,5471B 
was observed with a total exposure time of 30 min, while Shell~2 
only 10 min.  The exact slit positions, shown in 
Fig.~\ref{fig:ngc5471b}a, were recovered by matching the 
surface brightness profiles along the slits to those extracted 
from the WFPC2 \ha\ image.  The accuracy of the position matching 
is better than 0\farcs5.  The instrumental FWHM, determined from
Gaussian fits to the unresolved telluric lines, is 
$16.8 \pm 0.5$ \kms\ for NGC\,5471B, and $12.8 \pm 0.5$ \kms\ 
for Shell~2.

\begin{figure}
\plotone{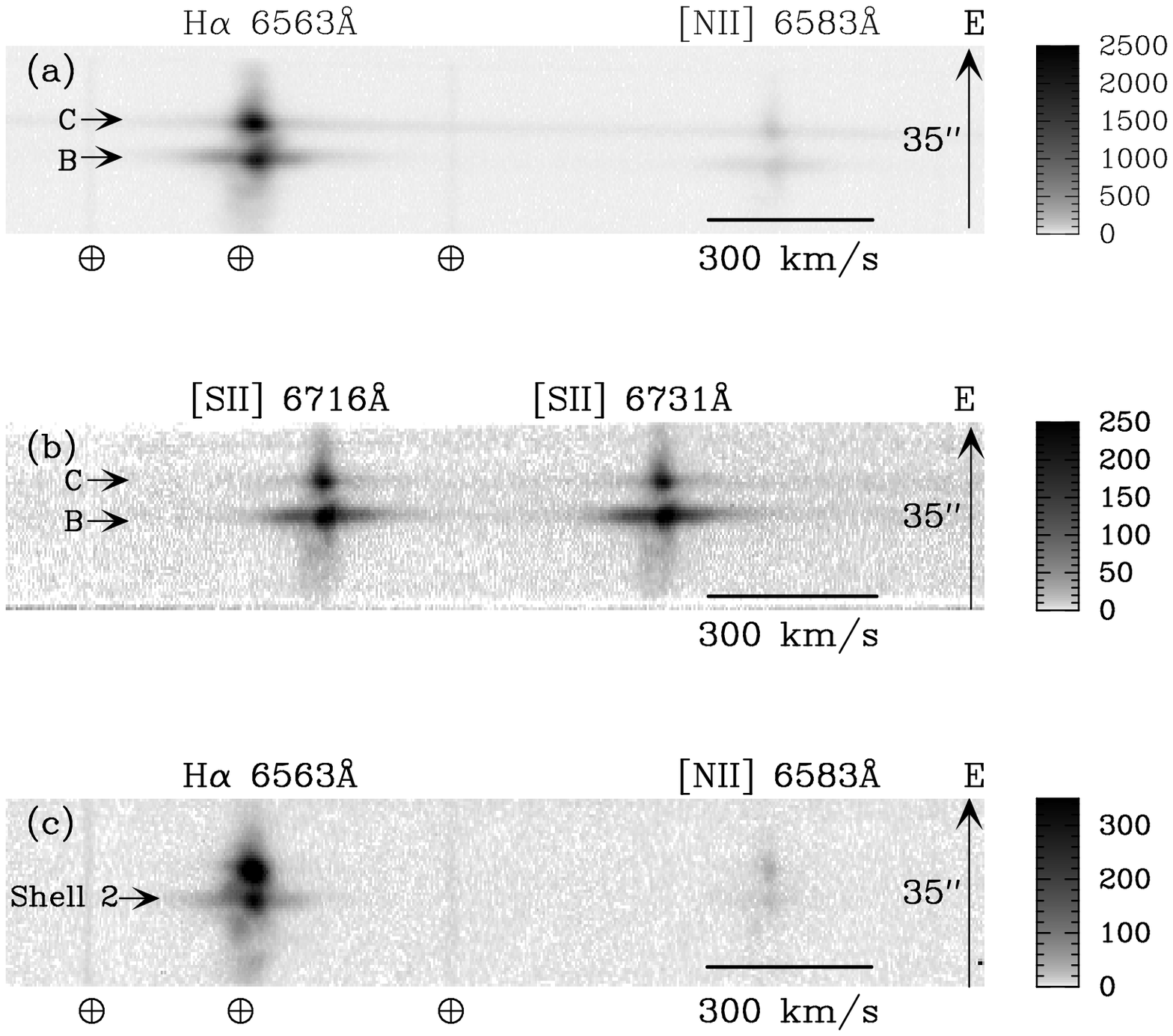}
\caption{KPNO 4~m echellograms of (a) the \ha+\nii\ $\lambda$6583 
lines, and (b) the \sii\ $\lambda\lambda$6716, 6731 lines along an 
E-W slit centered on NGC\,5471B (=Shell~1), and (c) the 
\ha+\nii\ $\lambda$6583 lines along an E-W slit centered on 
Shell~2.   The C-component of NGC\,5471, to the east of NGC\,5471B,
is also detected and marked in (a) and (b).  The bright source to the 
east of Shell 2 is a bright compact \hii\ region to the north of 
NGC\,5471C.
The images are displayed with square-root image transfer 
functions, as indicated by the greyscale bars on the right.  
The telluric lines marked by $\oplus$ are, from left to right, 
\ha\ $\lambda$6562.8, OH $\lambda$6568.8, and 
OH $\lambda\lambda$6577.2, 6577.4 lines.  
Both NGC\,5471B and Shell~2 have an average heliocentric velocity 
of $297\pm3$ \kms.}
\label{fig:echelle_raw}
\end{figure}

The \ha+\nii\ echellograms of NGC\,5471B and Shell~2 and the 
\sii\ $\lambda\lambda$6716, 6731 echellogram of NGC\,5471B are 
displayed in Figure~\ref{fig:echelle_raw}.
These new echelle data reveal high-velocity gas over a larger
velocity range than that reported by \citet{CK86} because of 
an improvement in the S/N of data.

The IRAF software was used for data reduction and analysis.  We 
followed the same reduction procedures as described in Paper~I.  
The sky-subtraction is more difficult for NGC\,5471B, because 
it is embedded in the giant \hii\ region NGC\,5471, and the nebular 
emission fills the entire length of the slit.  We have 
thus adopted the sky spectrum extracted from the observations of 
MF\,83 reported in Paper~I to remove the telluric lines in the
spectrum of NGC\,5471B.  The results are satisfactory, because 
MF\,83 is sufficiently close to NGC\,5471B and the observations 
of MF~83 were made immediately after those of NGC\,5471B.
The sky-subtracted \ha\ line profile of NGC\,5471B extracted 
over 2\farcs9 along the slit is Hanning smoothed over five adjacent 
pixels and displayed in Figure~\ref{fig:ha_fit}.

\section{Supernova Remnant Candidates in NGC\,5471}

SNRs are commonly diagnosed by their high \sii/\ha\ ratios,
as S$^+$ can be collisionally excited in the cooling 
region behind SNR shocks and emit strongly in the \sii\ 
$\lambda\lambda$6716, 6731 lines \citep{raymond79}.
Although the \sii/\ha\ ratios of known SNRs are frequently 
observed to be 0.5--1.0 \citep{Fe85}, optical SNR surveys have 
used a lower threshold, 0.4--0.45, to identify SNR candidates
because nebular backgrounds tend to lower the apparent
\sii/\ha\ ratios \citep[e.g.,][]{Lo90,MF97}.

The \sii/\ha\ ratio map of NGC\,5471 (Fig.~\ref{fig:ngc5471}f) 
shows three shells with enhanced \sii\ emission.
The positions, sizes, and \sii/\ha\ ratios of these three shells 
are listed in Table~\ref{tbl:shells}.
The apparent \sii/\ha\ 
ratios of these shells, 0.2--0.4, are higher than those of the
background \hii\ region, $\sim0.1$, but lower than the
SNR identification threshold.  These low observed ratios are not 
surprising, given that the background \hii\ region is extremely 
bright.  We have measured the average background emission from 
regions around the shells and subtracted the background from the 
\sii\ and \ha\ images.  The background-subtracted \sii/\ha\ 
ratios of the \sii-bright shells, also given in 
Table~\ref{tbl:shells}, are in the range 0.5--0.8, fully 
consistent with those of SNRs.

The \sii-bright shells in NGC\,5471 have sizes of 50--80 pc,
for a distance of 7.2 Mpc to M101 \citep{St98}.  
Interstellar shells with these large sizes and 
high \sii/\ha\ ratios can be either classical SNRs or superbubbles 
\citep{La77}. To distinguish between these two possibilities, 
the stellar content and nebular dynamics need to be 
examined.\footnote{The conventional SNR diagnostics, bright X-ray 
emission and nonthermal radio emission, are ineffective for SNRs 
in bright giant \hii\ regions such as NGC\,5471 because shocked
fast stellar winds can produce X-ray emission and the thermal
emission from the \hii\ region overwhelms the nonthermal emission
from SNRs.}
A shell is likely to be a superbubble, if it encompasses a high
concentration of blue stars and the shell has an expansion 
velocity $\ll$ 100 \kms.  Conversely, the existence of SNRs 
can be inferred, if no concentrations of blue stars are present
or the shell expansion velocity is high, $\gg$ 100 \kms.

Shell~1, the brightest of the three \sii-enhanced shells 
in NGC\,5471, corresponds to the SNR NGC\,5471B.
The $HST$ WFPC2 images in Fig.~2 show concentrations 
of blue stars within the boundary of Shell~1, making the
existence of a superbubble a distinct possibility.
On the other hand, the echelle spectrum of Shell~1 
(Fig.~\ref{fig:ha_fit}) shows high-velocity gas over a
velocity range of $\sim620$ \kms, which has been seen only 
in SNRs.  It is thus most likely
that Shell~1 contains a SNR but the physical conditions are 
complicated by the complex stellar environment.  A detailed 
analysis of Shell~1 will be presented in \S 4.

Shell~2 is at $\sim2''$ north of NGC\,5471B.  With a size 
of $52\times50$ pc, it is the smallest of the three shells.  
It exhibits a clear ring morphology in both the \ha\ and \sii\ 
lines.  The {\it F547M} continuum image (Fig.~\ref{fig:ngc5471b}c) 
shows sparse distribution of stars in the vicinity of Shell~2 
with no specific concentrations within the shell interior.  
The echellogram of Shell~2 (Fig.~\ref{fig:echelle_raw}c) 
shows a very broad \ha\ line with high-velocity gas detected 
over a velocity range of $\sim450$ \kms.   The lack of prominent 
concentrations of stars and the large expansion velocity in the 
high-velocity gas both support the existence of a SNR.  Even if 
Shell~2 is a superbubble formed by a small OB association, it 
must have been recently accelerated by a SNR.

Shell~3 is located at the low-density outskirts of NGC\,5471.
Its size, $\sim70$ pc, is comparable to that of Shell~1.
It has the highest \sii/\ha\ ratios among the three [S II]-bright
shells, and its background-subtracted \sii/\ha\ ratio is within the
higher range of SNRs.  The {\it F547M} continuum image 
(Fig.~\ref{fig:ngc5471}c) does not detect any bright stars
or concentrations of stars in the interior or along the 
rim of Shell~3.  It is unlikely that Shell~3 is a superbubble.
Although no kinematic information is available for Shell~3
and the size of Shell~3 is larger than the average size of
Galactic SNRs, we consider that Shell~3 is probably a bona fide
SNR and its large size is a direct consequence of the low density
in its interstellar environment.

\section{Analysis of Physical Properties of NGC\,5471B}

\subsection{Interstellar and Stellar Environment of NGC\,5471B}

As shown in Fig.~\ref{fig:ngc5471}, NGC\,5471B has a complex 
environment.  It is not trivial to separate the SNR from 
the background.  Different morphologies are seen in the \ha\ and
\sii\ images.   Since SNR shocks can be diagnosed by high \sii/\ha\ 
ratios, the \sii\ image should illustrate more clearly the boundary 
of the SNR.  Indeed, the continuum-subtracted \sii\ image provides 
the most well-defined limb-brightened shell morphology of 
NGC\,5471B.  The shell is elongated with a size of 
$2\farcs2 \times 1\farcs7$, or $77 \times 60$ pc.  A bright
emission patch is present in the southwest corner. 
Along this direction, \sii-bright emission is still present
exterior to the shell, making the SNR boundary somewhat ambiguous.
Owing to this patch of \sii-bright emission, the \sii/\ha\ 
ratio map in Fig.~2f shows high ratios over a region
$\sim50$\% larger than the \sii\ shell.  We will assume that 
the limb-brightened \sii\ shell represents the SNR in NGC\,5471B.

The continuum-subtracted \ha\ image shows an oval disk of emission 
with a slight limb-brightening at the west rim and a bright compact
emission region at the southeast corner. This bright compact region,
showing bright stellar emission in the continuum bands and having
the lowest \sii/\ha\ ratio, is most likely a high-excitation \hii\ 
region.  The oval disk of \ha\ emission, roughly coincident with the 
\sii\ shell, is probably associated with the SNR.  The SNR is 
superposed on a bright \hii\ background; the average surface 
brightness at the main body of the SNR is only 2-4 times as bright 
as those of the surrounding regions.

The {\it F547M} image shows many concentrations of stars in and 
around NGC\,5471B.  Within the main body of NGC\,5471B, two
peaks of stellar emission are present at the eastern end.
The brighter stellar source is coincident with the aforementioned 
bright high-excitation \hii\ region; the fainter stellar source 
is not associated with prominent \ha\ emission, but the \sii/\ha\ 
ratio in its vicinity is low, indicating that it is also inside 
an \hii\ region.  There is diffuse continuum emission along and 
interior to the south rim of the main body.  Since the {\it F547M}
band does not contain any bright nebular emission lines and 
the distribution of this continuum light is different from that 
of the \ha\ emission, this diffuse continuum emission most likely 
originates from an extended distribution of unresolved stars 
similar to the ``star clouds'' defined by \citet{LH70} for 
OB associations in the Large Magellanic Cloud (LMC).

Many peaks of stellar continuum emission are present in the
vicinity of NGC\,5471B (see Fig.~\ref{fig:stars}).
Their brightnesses in the {\it F547M}, 
{\it F675W}, and \ha-subtracted {\it F675W} bands ($m_{\it F547M}$,
$m_{\it F675W}$, and $m_{\it F675W'}$) are given in Table 2.
To determine the nature of these stellar sources, we present
in Figure~\ref{fig:cmd} a color-magnitude diagram of these 
sources using $m_{\it F547M}$ and $m_{\it F675W'}$.  
To facilitate comparisons with normal stars, 
we have synthesized $m_{\it F547M}$ and $m_{\it F675W}$ for stars 
of different spectral types and luminosity classes
using the \citet{Ku93} stellar atmosphere models and 
the WFPC2 filter transmission curves \citep{Bi96}.  
The synthetic photometry was calculated for a metallicity of
1/10~$Z_{\odot}$, a reddening of $E(B-V) = 0.16$ mag \citep{KG96},
and a distance modulus of 29.3 mag \citep[= 7.2 Mpc,][]{St98}.
In Fig.~\ref{fig:cmd}, the synthetic main sequence from O3\,V 
downward is plotted in a thick line and the supergiants in a 
thin line with spectral 
types marked.  It is immediately clear from the comparison that 
all identified stellar sources are either supergiants or
composite with multiple stars.  

While the brightest stellar source, \#18, is outside and to 
the south of the shell in NGC\,5471B, the next two brightest 
sources, \#21 and \#22, are the two stellar concentrations
inside the shell.  All these three bright sources have blue 
colors, and must be OB associations or clusters.  Source \#18,
not surrounded by a recognizable \hii\ region, must be the 
oldest among these three sources.  It is thus apparent that 
the SNR shell in NGC\,5471B encompasses two young OB associations. 

\subsection{Luminosity and Dynamics of the SNR Shell in NGC\,5471B}

\begin{figure}
\plotone{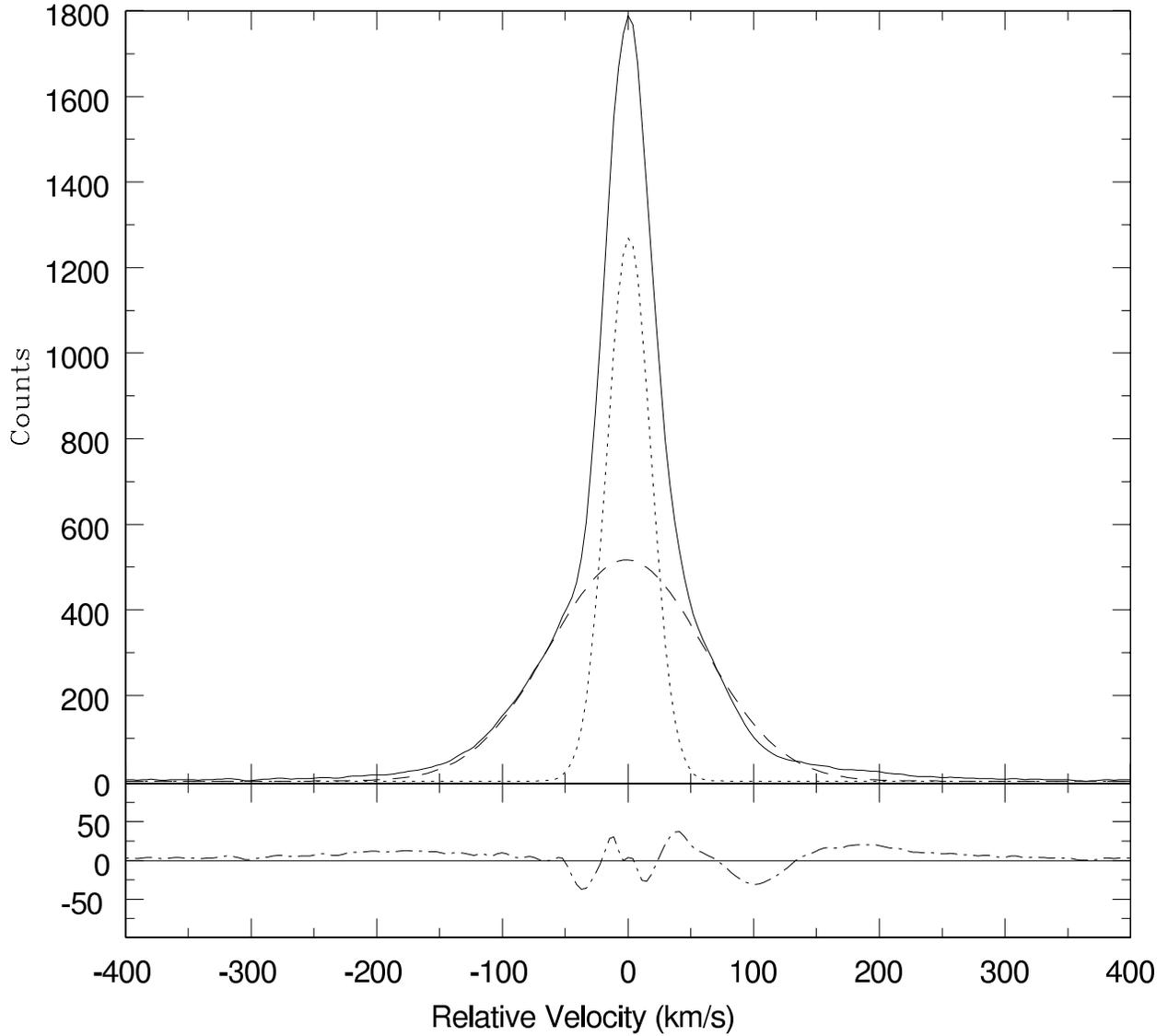}
\caption{\ha\ velocity profile of NGC\,5471B extracted from a
2$''$$\times$2\farcs9 echelle aperture centered on NGC\,5471B.
The profile (in solid line) is fitted by two Gaussian components
(in dotted and dashed lines).  The residual of the fit, displayed
in the lower panel, shows that high-velocity gas is detected
over more than 600 \kms.  The relative velocity in the horizontal 
axis is referenced to the heliocentric velocity of NGC\,5471B, 
297 \kms.}
\label{fig:ha_fit}
\end{figure}

\begin{figure}
\plotone{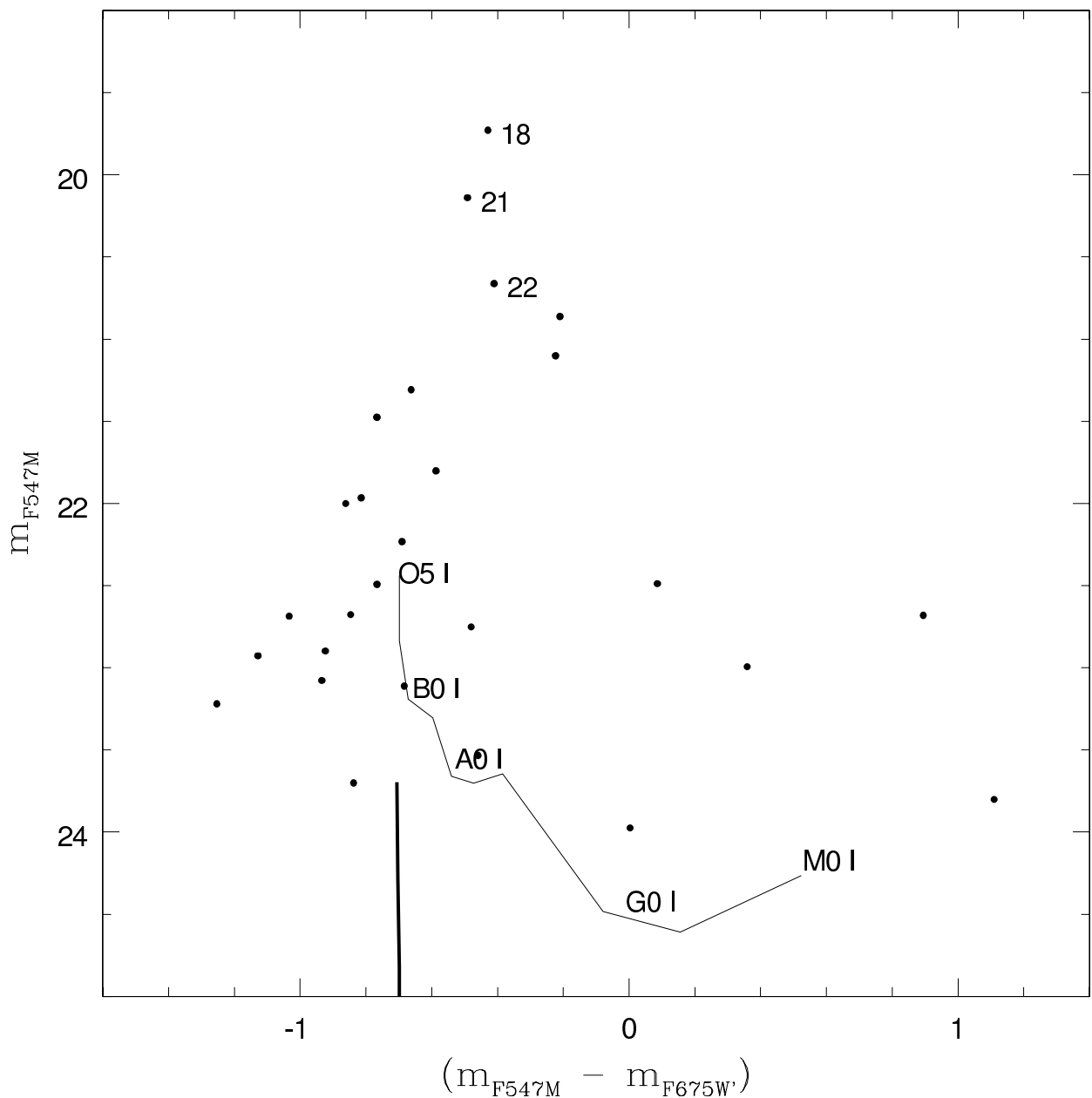}
\caption{Color-magnitude diagram of the stellar sources in
the vicinity of NGC\,5471B.  The stellar sources in Table 2
are plotted as $\bullet$.  The \ha-corrected {\it F675W} magnitude,
$m_{\it F675W'}$, is used to reduce photometric errors caused by \ha\ 
contamination in the {\it F675W} band.  Loci of the synthetic main 
sequence from O3\,V downward are plotted in a thick line and the supergiants 
in a thin line, for a metallicity of 1/10~$Z_{\odot}$, a reddening 
of $E(B-V) = 0.16$ mag, and a distance modulus of 29.3 mag.
The spectral types of the synthetic supergiants are marked.}
\label{fig:cmd}
\end{figure}

Several physical properties of the SNR shell in NGC\,5471B can be 
derived from the multi-order echelle observations.  First of all, 
the expansion of the shell can be determined from the \ha\ line 
profile.  The echelle image of the \ha\ and \nii\ lines of 
NGC\,5471B (Fig.~\ref{fig:echelle_raw}) shows narrow cores and 
extended wings superposed on a faint stellar continuum.
The presence of the extended wings in the \nii\ lines indicates that
this component originates from shocked gas, as opposed to stellar 
atmospheric emission from Wolf-Rayet (WR) stars or luminous blue 
variables (LBVs).  The \ha\ velocity profile of NGC\,5471B 
(Fig.~\ref{fig:ha_fit}), extracted over 2\farcs9 along the 2$''$-wide 
slit, can be fitted by a narrow component of FWHM = $41\pm2$ \kms\ 
and a broad component of FWHM = $148\pm5$ \kms.
The widths of the narrow and broad components are in the ranges typical 
for \hii\ regions and SNRs, respectively \citep{CK88,CK94b}. 
Therefore, we assign the narrow component to the background \hii\ 
region and the broad component to the SNR.

The decomposition of SNR and \hii\ region components within the
echelle aperture allows us to determine the \ha\ luminosity of the 
SNR from the $HST$ WFPC2 \ha\ image of NGC\,5471B.
The total \ha\ flux within the $2''\times2\farcs9$ aperture 
centered on the SNR shell is $(2.6\pm0.2)\times 10^{-13}$ 
ergs~cm$^{-2}$~s$^{-1}$.  The error originates mostly from the 
imprecise aperture of the echelle observation which is affected 
by the seeing.  The best spectral
fit to the \ha\ velocity profile indicates that the broad component is
1.4 times as bright as the narrow component.  Therefore, the \ha\ flux
from the SNR is $(1.5\pm0.1)\times 10^{-13}$ ergs~cm$^{-2}$~s$^{-1}$.
This flux is consistent with the value 
$1.5\times10^{-13}$~ergs~cm$^{-2}$~s$^{-1}$ estimated by
\citet{CK86}.  For a distance of 7.2 Mpc and a $c$(\hb) 
of 0.23 \citep{KG96}, the \ha\ luminosity of the SNR is 
$(1.4\pm0.1) \times 10^{39}$ ergs~s$^{-1}$.

The mass of the SNR can be derived from its \ha\ luminosity and
density.  Our echelle observation detected the 
\sii\ $\lambda\lambda$6716, 6731 doublet.  The velocity profiles
of the \sii\ doublets are likewise fitted by narrow and broad 
components.   The \sii\ $\lambda$6716/$\lambda$6731 ratio
is 1.4 and 1.0 for the narrow and broad components, respectively.
The \sii\ doublet ratio for the narrow component is at the 
low-density limit, implying that the electron density of the 
\hii\ region is $\ll50$~cm$^{-3}$.  The \sii\ doublet ratio
for the broad component, on the other hand, corresponds to
an electron density of $\sim700$~cm$^{-3}$, adopting an electron 
temperature of 13,000 K \citep{Sk85}.
This density and the above-derived \ha\ luminosity infer a mass of 
$4600\pm500$ M$_\odot$ in the SNR shell.  This mass is consistent with 
but more accurate than the value $6500\pm3000$ M$_\odot$ estimated by 
\citet{CK86} because of the updated atomic constants and more direct 
measurement of the electron density used in our current work.

Finally, the kinetic energy of the SNR shell can be estimated from
the mass and expansion velocity.
The expansion velocity of the SNR cannot be measured directly 
because the SNR is not spatially resolved in the echelle data.  
As the integrated velocity profile of the SNR is weighted by the 
surface brightness of each emission region within the SNR, the FWHM 
of the integrated profile cannot be used to determine the expansion
velocity of the SNR.  Instead, we adopt the observed extreme 
velocity offsets of the broad line as a lower limit of the expansion
velocity of the SNR.  The approaching and receding sides of the SNR 
are detected up to about $-$290 \kms\ and $+$330 \kms\ with respect 
to the systemic velocity of the \hii\ region, respectively (see 
Fig.~\ref{fig:ha_fit}).  Thus we adopt an expansion velocity of 
330 \kms\ for the SNR, and the kinetic energy of the SNR is 
$5.0 \times 10^{51}$ ergs.
The kinetic energy estimated by \citet{CK86} is 20 times smaller
because they used too small an expansion velocity for the SNR.

\subsection{X-ray Emission and Hot Ionized Gas of NGC\,5471B}

X-ray emission from NGC\,5471 has been detected by $ROSAT$ 
observations with both the Position Sensitive Proportional 
Counter (PSPC) and the High-Resolution Imager (HRI).
The angular resolutions of the PSPC and HRI were not adequate
to distinguish between point sources and diffuse emission from
NGC\,5471; however, the spectral shape of the PSPC spectrum
of NGC\,5471 is consistent with that of thin plasma emission
\citep{WC95}.  Using an ultradeep, 227 ks, $ROSAT$ HRI 
image of M101, \citet{WIP99} detected NGC\,5471 at a count
rate of 1.22 counts ks$^{-1}$, corresponding to an X-ray
luminosity of $3\times10^{38}$ ergs~s$^{-1}$ \citep{Wa99}.

\begin{figure}
\plotone{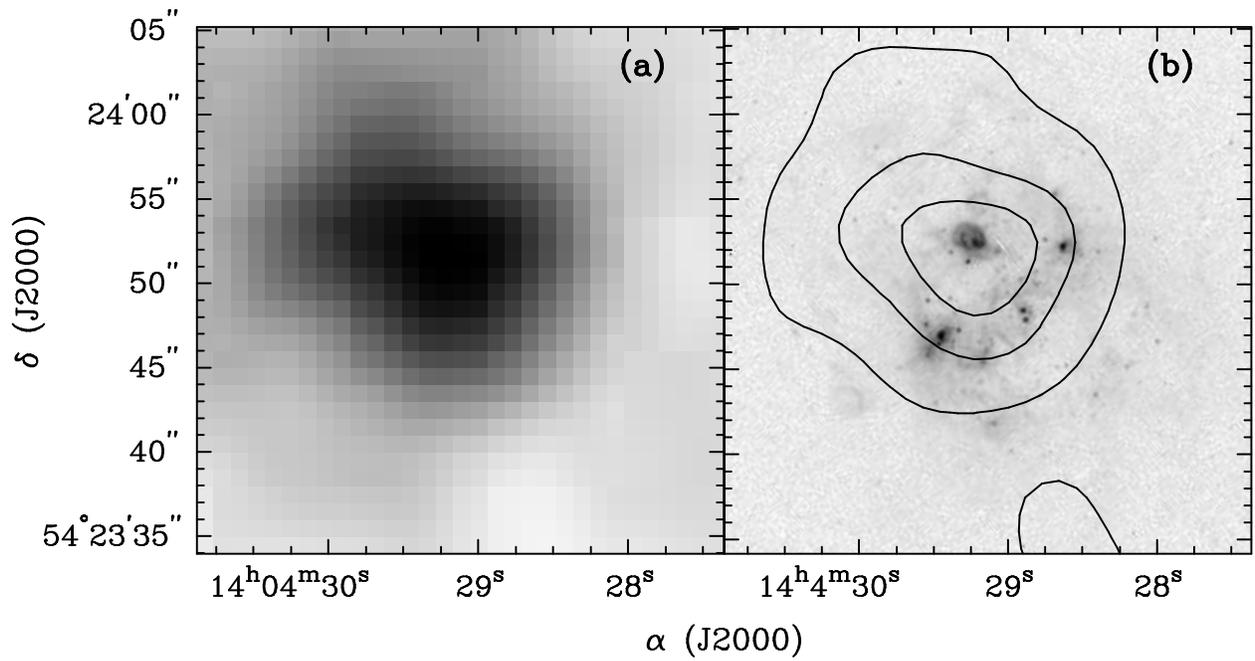}
\caption{(a) $ROSAT$ HRI image of NGC\,5471.  This HRI image 
is extracted from a 227 ks observation, and has been smoothed 
with a Gaussian of $\sigma = 3''$. (b) $HST$ WFPC2 \sii\ image
overlaid with the HRI X-ray contours at levels of 25\%, 50\%, 
75\% and 90\% of the peak value.}
\label{fig:xray}
\end{figure}

We have plotted the X-ray contours extracted from the 227 ks
$ROSAT$ HRI image over the $HST$ WFPC2 \sii\ image of NGC\,5471
in Figure~\ref{fig:xray}.  The X-ray contours are centered near
NGC\,5471B.  Given the combined pointing errors of the $HST$
and $ROSAT$, a few arcsec, it is very likely that the X-ray
source is centered at NGC\,5471B.
The high-resolution $Chandra$ observation of NGC\,5471
(scheduled for Cycle 3) will be able to determine unambiguously
the position and spatial extent of this X-ray emission.

\citet{Wa99} used the observed X-ray emission from NGC\,5471 
in conjunction with the Sedov solution for a SNR, and computed
a thermal energy of $\sim10^{52}$ ergs for the hot gas in the
NGC\,5471B SNR.  This amount of thermal energy is about twice
as high as the kinetic energy we have derived for the 10$^4$ K
ionized expanding shell of this SNR.

\section{Discussion}

The term ``hypernova" has been vaguely used in theoretical
contexts until recently when the connection between GRBs and 
supernovae was illustrated observationally through SN\,1988bw 
\citep{Ketal98} and SN\,1997cy \citep{Getal00}.  
Analyses of the light curves and spectral variations of these
two supernovae indicate that the kinetic energy released in the 
supernova explosion was $3-5\times10^{52}$ ergs \citep{Netal01,
Tetal00}.  Such explosion energies are more than an order of magnitude
higher than the canonical supernova explosion energy of 10$^{51}$ 
ergs, qualifying these supernovae as ``hypernovae".  
Hypernovae without GRB counterparts have also been reported, e.g.,
SN\,1997ef \citep{MIN00}.  
Thus, we consider a supernova with an explosion energy greater
than 10$^{52}$ ergs as a hypernova, and a SNR that requires such
a high explosion energy as a hypernova remnant.

The SNR in NGC\,5471B has been suggested to be a hypernova remnant
because its X-ray emission implies a thermal energy that requires
a supernova explosion energy greater than 10$^{52}$ ergs \citep{Wa99}.
Our high-resolution imaging and spectroscopic observations at optical
wavelengths have allowed us to (1) resolve the shell structure of the 
SNR, (2) examine the stellar and interstellar environments of the SNR,
(3) study the expansion dynamics of the SNR, (4) separate the \ha\
flux of the SNR from the background \hii\ region emission, and (5) 
determine the mass and kinetic energy of the SNR shell.
Using these results we are now able to critically examine the 
nature of the hypernova remnant candidate NGC\,5471B.

\subsection{Explosion Energy Requirement of the SNR Shell in NGC\,5471B}

NGC\,5471B possesses typical signatures of SNRs, e.g., nonthermal
radio emission, high \sii/\ha\ ratio, and X-ray spectrum consistent
with those of thin plasma emission \citep{Sk85,WC95}.  It is most
certain that NGC\,5471B contains a SNR.  The $HST$ WFPC2 images of 
NGC\,5471B show a dominant, \sii-bright $77\times60$ pc shell, which 
is responsible for the enhanced \sii/\ha\ ratio.
The echelle observations of NGC\,5471B have established that there
is a broad velocity component associated with high-velocity shocks.
From the comparison between the \sii\ and \ha\ velocity profiles 
of NGC\,5471B (Fig.~\ref{fig:echelle_raw}), it is clear that 
the broad velocity component is responsible for the \sii-enhancement.  
Therefore, the \sii-bright shell shown in the WFPC2 images must 
produce the broad velocity component in the echelle spectra.  

We can further confirm that the \sii-bright shell corresponds to the 
broad velocity component using a consideration of the brightnesses of 
the shell and the background \hii\ region.  The \ha\ velocity profile 
shows that the broad component contributes to 1.4/(1+1.4) = 58\% of 
the total \ha\ flux within the $2''\times2\farcs9$ echelle aperture.  
Using the WFPC2 \ha\ image and assuming that the shell is superposed 
on an \hii\ region with an average surface brightness similar to 
that of the surrounding regions, we find that the shell contributes to
$\sim54$\% of the total \ha\ flux within a $2''\times2\farcs9$
aperture centered on the shell. 
The similarity in the shell's and the broad component's contributions
to the total flux within the same aperture strengthens the conclusion 
that the \sii-bright shell gives rise to the broad velocity component.
The shell expansion velocity, $\sim330$ \kms, is higher than those of 
most SNRs in the LMC \citep{CK88}, and certainly can produce the radio 
and X-ray signatures of SNRs by shocks\footnote{Note that the shock 
velocities are higher than the observed expansion velocity of the optical 
SNR shell.  For example, the velocity of an adiabatic shock is 4/3 times 
the shell expansion velocity.  Furthermore, the optically detected 
material is likely associated with SNR shocks propagating through dense 
clouds, while the shocks in the intercloud medium are faster but their 
post-shock material is too tenuous to be detected.}.
Therefore, we believe that all observed SNR signatures originate from
this \sii-bright shell, and will identify this shell as the SNR.

As we have derived in \S4.2, the expanding SNR shell in NGC\,5471B 
has a mass of 4,600~M$_\odot$ and a kinetic energy of
$5.0\times10^{51}$ ergs.  
The kinetic energy in a mature SNR shell is much lower than 30\% of 
the supernova explosion energy \citep{Ch74}; thus the observed kinetic 
energy in the SNR shell in NGC\,5471B requires an explosion energy 
greater than 10$^{52}$~ergs.
Moreover, \citet{Wa99} assumed that the X-ray emission from NGC\,5471B
originates from the hot shocked gas in a SNR, and derived a thermal 
energy of $\sim10^{52}$~ergs for the SNR.
Both the optical and X-ray considerations require a supernova explosion 
energy $>10^{52}$ ergs; therefore, the \sii-bright expanding shell in
NGC\,5471B must be a hypernova remnant or energized by a burst of
multiple supernova explosions.

\subsection{Comparison with Other SNRs in Nearby Star Formation
Regions}

Was the \sii-bright shell in NGC\,5471B produced by a single hypernova 
explosion?  It is almost impossible to answer this question, especially 
since the shell is in an active star formation region with two prominent 
OB associations (see Figs.~3 \& 6) projected within the shell, as
described in \S4.1.
It is nevertheless possible to compare NGC\,5471B to other SNRs in 
nearby active star formation regions in order to determine whether 
the shell in NGC\,5471B is a common entity.  We will use the sample 
of star formation regions in the LMC and NGC\,604 in M33 for 
comparison.

Convincing SNR signatures have been detected in at least 13 OB 
associations in the LMC \citep{Ch97}.  Of these, eight contain 
SNRs showing signatures at all wavelengths, and all of them are
in the outskirts of their associated OB associations.
The other five contain SNRs within their superbubbles and show only
X-ray signatures.   These superbubbles have sizes 50--100 pc, 
comparable to that of the shell in NGC\,5471B; however, their 
expansion velocities are $\ll100$ \kms, much slower than that of 
the NGC\,5471B shell.  The shell in NGC\,5471B is much more
energetic than the normal superbubbles around OB associations. 

We may further dismiss the two prominent OB associations projected 
within the shell of NGC\,5471B as the providers of multiple supernovae.
The existence of bright compact \hii\ regions around these OB associations
indicate that they are young and the surrounding interstellar gas has not
been dispersed by SNR shocks.  It is thus unlikely that the two OB 
associations in the NGC\,5471B shell energized the SNR shell.
Moreover, the size and expansion velocity of the SNR shell implies an
upper limit of $1\times10^5$ yr for the dynamic age.  
In order to have more than 10 normal supernovae from an OB association
within such a small time span, these supernovae must have the 
shortest-lived progenitors, i.e., the most massive stars.  
Assuming that all 10 progenitors were O3\,V stars with initial
masses of 100--120 M$_\odot$ and adopting Salpeter's (1955)
slope for the initial mass function, we expect the OB associations 
to contain 250 stars in the mass range of 25--100 M$_\odot$, 
corresponding to the later-type O stars \citep{Sc97}.
The $m_{\it F547M}$ magnitudes of the two OB associations 
projected within the NGC\,5471B shell are equivalent to
70 and 45 O6\,V stars for sources \#21 and \#22, respectively.
The total number of O stars in these two OB associations is most
likely smaller than 115 because some of the stars may be supergiants
that are several times brighter than their main-sequence counterparts.
Therefore, the expected number of O stars exceeds the total
number of O stars implied by $m_{\it F547M}$ by more than a
factor of 2.  Thus, the observed OB associations in the shell 
of NGC\,5471B are unable to produce enough supernovae within 
the dynamic age and power the shell.

\begin{figure}
\plotone{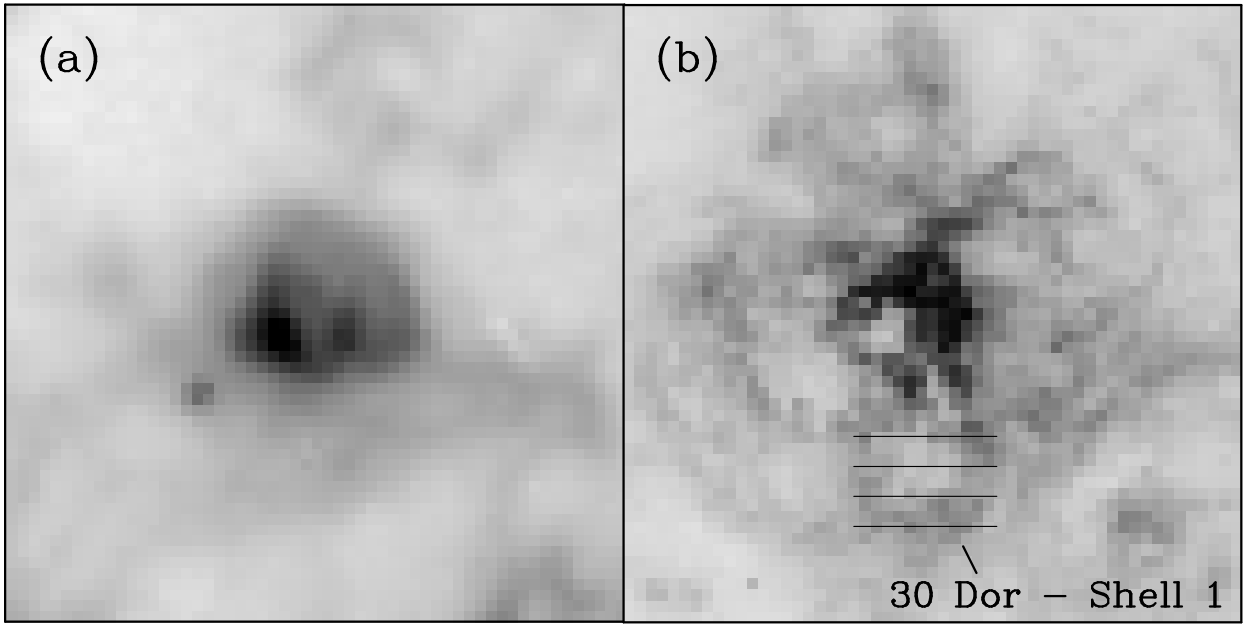}
\caption{(a) $HST$ WFPC2 \ha\ image of NGC\,5471B.
(b) Ground-based \ha\ image of 30~Dor binned to simulate 
a $HST$ WFPC2 image of 30~dor at the distance of M101.
Both images have a linear dimension of 200 pc $\times$ 
200 pc, and are displayed with square-root image transfer 
functions.  30 Dor-Shell~1 and the echelle slit positions 
from Chu \& Kennicutt (1994a) are marked over 30 Dor in (b).}
\label{fig:com_im}
\end{figure}

It would be appropriate to compare NGC\,5471B to the giant
\hii\ region 30 Doradus in the LMC, as 30 Dor is the most active
star formation and has the highest concentration of massive stars
in the LMC.  Figure~\ref{fig:com_im} displays NGC\,5471B and
30 Dor at the same physical scale over an area of 200 pc $\times$ 
200 pc, adopting a distance of 50 kpc to the LMC \citep{Fe99}.  
Within this region, the \ha\ luminosity of NGC\,5471B is 
$4.6\times10^{39}$ ergs~s$^{-1}$, and 30 Dor $6.2\times10^{39}$ 
ergs~s$^{-1}$ \citep{KH86,MCP85}.
The similarity in the \ha\ luminosity implies a similarity in the 
ionizing power of stars encompassed in these two regions.  It would 
then be truly a fair comparison between NGC\,5471B and 30 Dor.

\begin{figure}
\plotone{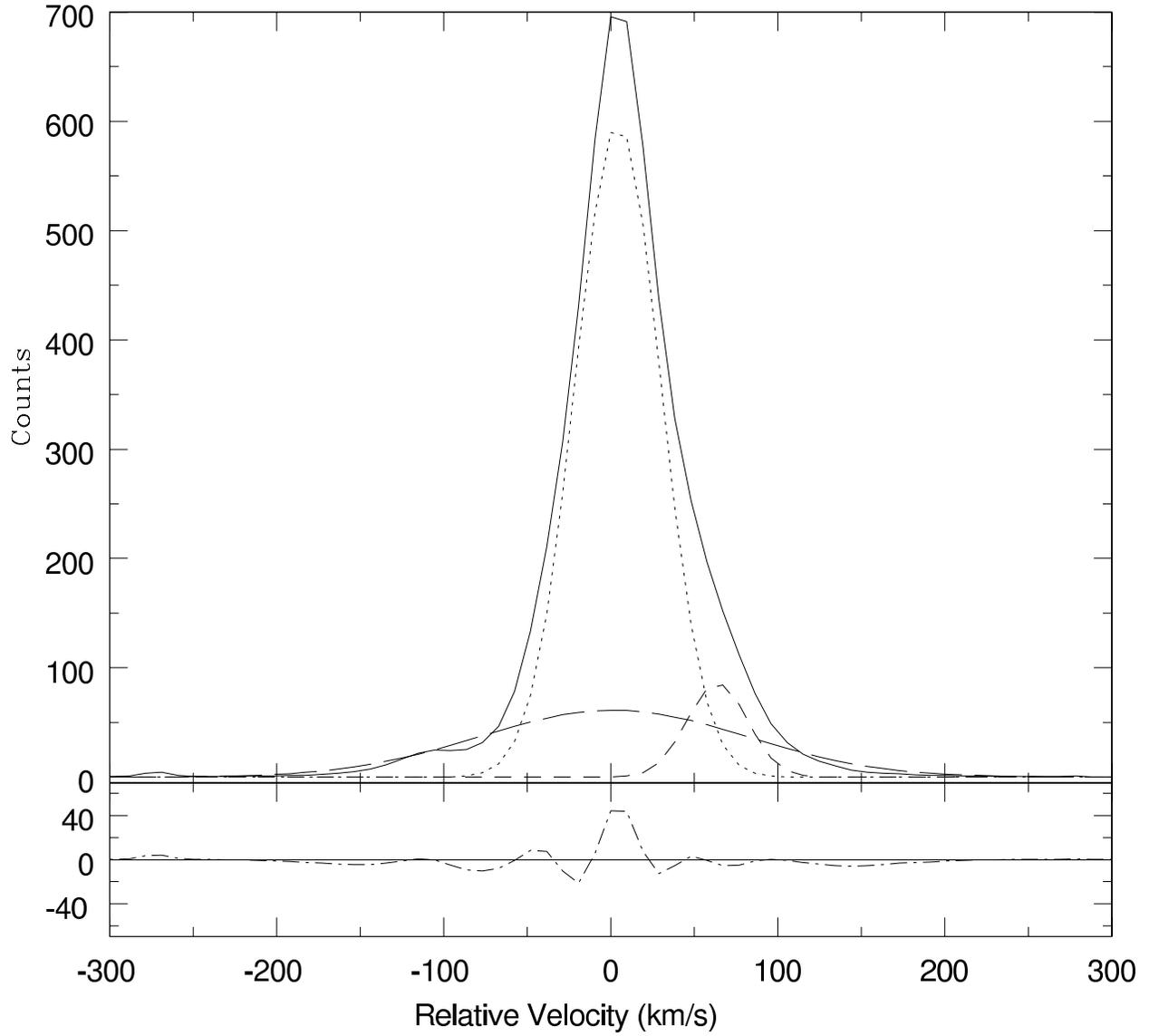}
\caption{Pseudo-integrated \ha\ velocity profile of Shell~1 
in 30 Dor.  The profile (in solid line) is fitted by three
Gaussian components (in dotted, short-dashed, and long-dashed
lines).  The residual of the fit is displayed in the lower panel.
The relative velocity in the horizontal axis is referenced to the 
heliocentric velocity of 30 Dor, 275 \kms.}
\label{fig:30dor_ech}
\end{figure}

Despite the similarity in ionizing power, the nebular structure 
of NGC\,5471B is very different from 30 Dor.  NGC\,5471B is
dominated by the SNR shell, while 30 Dor is dominated by several
shell structures of comparable sizes.  The
kinematic structure of 30 Dor has been mapped in great detail 
by \citet{CK94a}.  The central shell around the R136 cluster in 
30 Dor is much smaller (30 pc in diameter) and expands much more 
slowly ($\sim130$ \kms) than the shell in NGC\,5471B.
The five largest shells in 30 Dor, designated as Shells 1--5 by 
\citet{WF91}, show a common kinematic structure: a slowly expanding 
shell with V$_{\rm exp} \ll$ 100 \kms\ superposed by additional 
shocked clouds with velocity offsets up to $\pm300$ \kms.  
The \ha\ fluxes from these shells in 30 Dor are
dominated by the slow expanding shell component.  We use Shell 1
in 30 Dor as an example.  The echelle observations along four slit 
positions in 30 Dor-Shell 1, marked in Fig.~\ref{fig:com_im}, are 
used to simulate an integrated \ha\ velocity profile of 
30 Dor-Shell 1, shown in Figure~\ref{fig:30dor_ech}.  The 
``pseudo-integrated" profile of 30 Dor-Shell 1 shows a narrow core 
and broad wings, but the broad wings are highly asymmetric and the 
broad component contributes to only 22\% of the total flux. 

The fast expanding shell in NGC\,5471B is unmatched by any 
comparable shells in the LMC.  The giant \hii\ region NGC\,604
in M33 has also been studied in detail with similar echelle 
observations \citep{YCST96}.  Its kinematic structure is very
similar to that of 30 Dor, and it does not have any shell with an
expansion velocity or luminosity matching those of the shell in 
NGC\,5471B.  Therefore, it becomes clear that the fast expanding
shell in NGC\,5471B is truly outstanding and unique.  It could 
not have been produced by multiple supernovae as in the giant
\hii\ regions 30 Dor and NGC\,604.  Therefore, we conclude that
it is very likely that the energetic SNR in NGC\,5471B was produced
by a hypernova.

\begin{acknowledgments}

This research is supported by the NASA grant STI6829.01-95A.

\end{acknowledgments}

\clearpage

\clearpage

\begin{table}[h]
\caption{Journal of $HST$ WFPC2 Observations of NGC\,5471}
\label{tbl:log}
\begin{tabular}{lcrcll}
\tableline
\tableline

Filter	& $\lambda$\tablenotemark{a} & $\Delta\lambda$\tablenotemark{a}     
&Band	& Exposure	&Image Root Name\\
	& (\AA )  & (\AA) & &~~~(sec)	&		\\
\tableline

F656N	& 6562  & 22	& \ha\             & ~1 $\times$ 180   &u4dn0305r    \\
	&	&	&		  & ~2 $\times$ 600   &u4dn0306r, u4dn0307r \\	
F673N 	& 6733  & 47	& \sii\     & ~3 $\times$ 700 &u4dn0301r, u4dn0302r, u4dn0303r\\
F547M	& 5454  & 487 	&Str\"{o}mgren $y$& ~2 $\times$ 100   &u4dn030fr, u4dn030gr \\
	&	&	&		  & ~2 $\times$ 600   &u4dn030dr, u4dn030er \\ 
F675W	& 6997  & 889	& WFPC2 $R$       & ~2 $\times$ ~50   &u4dn030ar, u4dn030br \\
	&	&	&		  & ~2 $\times$ 400   &u4dn0308r, u4dn0309r \\
\tableline
\tablenotetext{a}{The mean wavelengths ($\lambda$) and widths ($\Delta\lambda$)
  of the filters are taken from the WFPC2 Instrument Handbook \citep{Bi96}.}
\end{tabular}
\end{table}

\begin{table}[h]
\caption{Photometry of Stellar Sources in NGC\,5471B}
\label{tbl:photos}
\begin{tabular}{rcccccr}
\tableline
\tableline

ID & R.A.	& Decl.	    & $m_{\it F547M}$ & $m_{\it F675W}$  &
$m_{\it F675W'}\tablenotemark{a}$ &$m_{\it F547M}-m_{\it F675W'}$\\
 & (J2000)	& (J2000) & (mag) & (mag) & 
(mag) & (mag)~~~~ \\
\tableline

 1 &14 04 28.94 &54 23 53.9 &22.68 $\pm$ 0.08&24.11 $\pm$ 0.25&23.52 $\pm$ 0.12&$-$0.85 $\pm$ 0.14 \\
 2 &14 04 29.00 &54 23 53.0 &22.23 $\pm$ 0.10&23.15 $\pm$ 0.20&22.92 $\pm$ 0.13&$-$0.69 $\pm$ 0.16 \\
 3 &14 04 29.02 &54 23 53.6 &21.48 $\pm$ 0.04&22.36 $\pm$ 0.05&22.24 $\pm$ 0.05&$-$0.77 $\pm$ 0.06 \\
 4 &14 04 29.03 &54 23 50.9 &23.80 $\pm$ 0.47&23.58 $\pm$ 0.56&22.69 $\pm$ 0.11& ~1.11 $\pm$ 0.48 \\
 5 &14 04 29.04 &54 23 50.2 &22.99 $\pm$ 0.21&21.91 $\pm$ 0.10&22.64 $\pm$ 0.10& ~0.36 $\pm$ 0.23 \\
 6 &14 04 29.06 &54 23 50.7 &22.49 $\pm$ 0.10&22.49 $\pm$ 0.15&22.40 $\pm$ 0.08& ~0.09 $\pm$ 0.13 \\
 7 &14 04 29.07 &54 23 51.8 &20.86 $\pm$ 0.03&20.88 $\pm$ 0.05&21.07 $\pm$ 0.03&$-$0.21 $\pm$ 0.05 \\
 8 &14 04 29.09 &54 23 53.8 &21.97 $\pm$ 0.06&22.65 $\pm$ 0.08&22.78 $\pm$ 0.10&$-$0.81 $\pm$ 0.12 \\
 9 &14 04 29.10 &54 23 53.4 &21.80 $\pm$ 0.07&22.51 $\pm$ 0.09&22.39 $\pm$ 0.08&$-$0.59 $\pm$ 0.10 \\
10 &14 04 29.10 &54 23 54.3 &23.08 $\pm$ 0.19&23.55 $\pm$ 0.23&24.01 $\pm$ 0.28&$-$0.93 $\pm$ 0.34 \\
11 &14 04 29.10 &54 23 54.8 &22.90 $\pm$ 0.09&24.62 $\pm$ 0.56&23.82 $\pm$ 0.17&$-$0.92 $\pm$ 0.20 \\
12 &14 04 29.12 &54 23 50.1 &23.53 $\pm$ 0.31&23.29 $\pm$ 0.32&23.99 $\pm$ 0.36&$-$0.46 $\pm$ 0.48 \\
13 &14 04 29.12 &54 23 51.4 &22.75 $\pm$ 0.18&23.71 $\pm$ 0.86&23.23 $\pm$ 0.36&$-$0.48 $\pm$ 0.41 \\
14 &14 04 29.15 &54 23 53.3 &21.47 $\pm$ 0.05&22.21 $\pm$ 0.09&22.24 $\pm$ 0.08&$-$0.77 $\pm$ 0.09 \\
15 &14 04 29.16 &54 23 54.2 &22.00 $\pm$ 0.05&22.71 $\pm$ 0.09&22.86 $\pm$ 0.08&$-$0.86 $\pm$ 0.09 \\
16 &14 04 29.18 &54 23 51.4 &23.97 $\pm$ 0.81&23.31 $\pm$ 0.85&23.97 $\pm$ 0.83& ~0.00 $\pm$ 1.16 \\
17 &14 04 29.21 &54 23 53.1 &22.68 $\pm$ 0.26&20.94 $\pm$ 0.08&21.79 $\pm$ 0.11& ~0.89 $\pm$ 0.28 \\
18 &14 04 29.27 &54 23 51.3 &19.73 $\pm$ 0.01&20.09 $\pm$ 0.02&20.16 $\pm$ 0.01&$-$0.43 $\pm$ 0.01 \\
19 &14 04 29.28 &54 23 50.0 &23.22 $\pm$ 0.11&24.24 $\pm$ 0.28&24.47 $\pm$ 0.25&$-$1.25 $\pm$ 0.27 \\
20 &14 04 29.28 &54 23 54.5 &22.49 $\pm$ 0.06&22.96 $\pm$ 0.16&23.26 $\pm$ 0.13&$-$0.77 $\pm$ 0.14 \\
21 &14 04 29.29 &54 23 52.4 &20.14 $\pm$ 0.02&19.66 $\pm$ 0.03&20.63 $\pm$ 0.04&$-$0.49 $\pm$ 0.05 \\
22 &14 04 29.29 &54 23 52.9 &20.66 $\pm$ 0.04&20.32 $\pm$ 0.07&21.07 $\pm$ 0.07&$-$0.41 $\pm$ 0.08 \\
23 &14 04 29.33 &54 23 50.2 &22.69 $\pm$ 0.07&23.82 $\pm$ 0.27&23.72 $\pm$ 0.14&$-$1.03 $\pm$ 0.16 \\
24 &14 04 29.34 &54 23 53.8 &23.11 $\pm$ 0.17&23.40 $\pm$ 0.21&23.79 $\pm$ 0.20&$-$0.68 $\pm$ 0.26 \\
25 &14 04 29.34 &54 23 54.3 &22.93 $\pm$ 0.10&24.05 $\pm$ 0.36&24.06 $\pm$ 0.23&$-$1.13 $\pm$ 0.25 \\
26 &14 04 29.39 &54 23 51.8 &21.10 $\pm$ 0.03&20.89 $\pm$ 0.03&21.32 $\pm$ 0.03&$-$0.22 $\pm$ 0.04 \\
27 &14 04 29.44 &54 23 53.1 &21.31 $\pm$ 0.03&21.84 $\pm$ 0.06&21.97 $\pm$ 0.04&$-$0.66 $\pm$ 0.05 \\
28 &14 04 29.46 &54 23 51.7 &23.70 $\pm$ 0.19&23.59 $\pm$ 0.24&24.54 $\pm$ 0.38&$-$0.84 $\pm$ 0.43 \\

\tableline
\tablecomments{Units of right ascension are hours, minutes, and seconds, and units
  of declination are degrees, arminutes, and arcseconds.}
\tablenotetext{a}{Apparent magnitudes measured from the \ha-subtracted {\it F675W}
 image of NGC\,5471B.}
\end{tabular}
\end{table}

\begin{table}[h]
\caption{\sii -Bright Shells in NGC\,5471}
\label{tbl:shells}
\begin{tabular}{ccccccc}
\tableline
\tableline

Shell 	&  R.A.  &  Decl.  & Angular Size & Linear Size\tablenotemark{a}& Apparent
  & Corrected\tablenotemark{b} \\
ID	& (J2000)  &  (J2000)  &  ($''$)       & (pc)	& \sii/\ha\ &  \sii/\ha\ \\

\tableline
1 & 14 04 29.27 & 54 23 52.5 & 2.2 $\times$ 1.7 & 77 $\times$ 60& 0.33 $\pm$ 0.03 & 0.5 $\pm$ 0.1\\
2 & 14 04 29.20 & 54 23 54.7 & 1.5 $\times$ 1.4 & 52 $\times$ 50& 0.23 $\pm$ 0.03 & 0.5 $\pm$ 0.2\\ 
3 & 14 04 30.08 & 54 23 43.0 & 2.0 $\times$ 1.9 & 70 $\times$ 67& 0.40 $\pm$ 0.05 & 0.8 $\pm$ 0.2\\

\tableline
\tablecomments{Units of right ascension are hours, minutes, and seconds, and units
  of declination are degrees, arminutes, and arcseconds.}
\tablenotetext{a}{For a distance of 7.2 Mpc to M101 \citep{St98}.}
\tablenotetext{b}{These corrected ratios are derived with an average background 
 \hii\ region emission subtracted.}
\end{tabular}
\end{table}


\begin{thebibliography}{}
\bibitem[Biretta et al.(1996)]{Bi96} Biretta, J.\ A., et al.\ 1996, 
 WFPC2 Instrument Handbook, Version 4.0 (Baltimore: STScI)
\bibitem[Chen et al.(2000)]{Ch00} Chen, C.-H.\ R., Chu, Y.-H., 
 Gruendl, R.\ A., \& Points, S.\ D.\ 2000, \aj, 119, 1317
\bibitem[Chevalier(1974)]{Ch74} Chevalier, R.\ A.\ 1974, \apj, 188, 501
\bibitem[Chu(1997)]{Ch97} Chu, Y.-H.\ 1997, \aj, 113, 1815
\bibitem[Chu \& Kennicutt(1986)]{CK86} Chu, Y.-H., \& Kennicutt, R.\ 
 C., Jr.\ 1986, \apj, 311, 85
\bibitem[Chu \& Kennicutt(1988)]{CK88} Chu, Y.-H., \& Kennicutt, R.\ 
 C., Jr.\ 1988, \aj, 95, 1111
\bibitem[Chu \& Kennicutt(1994a)]{CK94a} Chu, Y.-H., \& Kennicutt, R.\ 
 C., Jr.\ 1994a, \apj, 425, 720
\bibitem[Chu \& Kennicutt(1994b)]{CK94b} Chu, Y.-H., \& Kennicutt, R.\ 
 C., Jr.\ 1994b, ApSS, 216, 253
\bibitem[Feast(1999)]{Fe99} Feast, M.\ 1999, IAU Symp.~190: New Views 
 of the Magellanic Clouds, 190, 542
\bibitem[Fesen, Blair, \& Kirshner(1985)]{Fe85} Fesen, R.\ A., Blair, 
 W.\ P., \& Kirshner, R.\ P.\ 1985, \apj, 292, 29
\bibitem[Fryer \& Woosley(1998)]{FW98} Fryer, C., \& Woosley, S.\
  1998, \apjl, 502, 9
\bibitem[Germany et al.(2000)]{Getal00} Germany, L.~M., Reiss, D.~J., 
  Sadler, E.~M., Schmidt, B.~P., \& Stubbs, C.~W.\ 2000, \apj, 533, 320
\bibitem[Holtzman et al.(1995)]{Ho95} Holtzman, J.\ A., et al.\ 1995, 
 \pasp, 107, 1065
\bibitem[Jones et al.(1998)]{Jo98} Jones, T.\ W., et al.\ 1998, \pasp, 
 110, 125
\bibitem[Kennicutt \& Garnett(1996)]{KG96} Kennicutt, R.\ C., \& 
 Garnett, D.\ R.\ 1996, \apj, 456, 504
\bibitem[Kennicutt \& Hodge(1986)]{KH86} Kennicutt, R.\ C., \&
 Hodge, P.\ 1986, \apj, 306, 130
\bibitem[Kulkarni et al.(1998)] {Ketal98} Kulkarni, S.\ R.\ et al.\
 1998, Nature, 395, 663
\bibitem[Kurucz(1993)]{Ku93} Kurucz, R.\ 1993, ATLAS9 
 Stellar Atmosphere Programs and 2 km/s grid.\ Kurucz CD-ROM No.\ 
 13.\ Cambridge, Mass.: Smithsonian Astrophysical Observatory, 1993., 
 13
\bibitem[Lai et al.(2001)]{La01} Lai, S.-P., Chu, Y.-H., Chen, 
 C.-H.\ R., Ciardullo, R., \& Grebel, E.\ K.\ 2001, \apj, 547, 754 
 (Paper~I)
\bibitem[Lasker(1977)]{La77} Lasker, B.\ M.\ 1977, \apj, 212, 390
\bibitem[Long et al.(1990)]{Lo90} Long, K.\ S., Blair, W.\ P., 
 Kirshner, R.\ P., \& Winkler, P.\ F.\ 1990, \apjs, 72, 61
\bibitem[Lucke \& Hodge(1970)]{LH70} Lucke, P.\ B., \&
  Hodge, P.\ W.\ 1970, AJ, 75, 171
\bibitem[Mathis, Chu, \& Peterson(1985)]{MCP85} Mathis, J.~S., 
 Chu, Y.-H., \& Peterson, D.~E.\ 1985, \apj, 292, 155
\bibitem[Matonick \& Fesen(1997)]{MF97} Matonick, D.\ M., \& Fesen, 
 R.\ A.\ 1997, \apjs, 112, 49
\bibitem[Mazzali, Iwamoto, \& Nomoto(2000)]{MIN00}  Mazzali, P.~A., 
  Iwamoto, K., \& Nomoto, K.\ 2000, \apj, 545, 407
\bibitem[Nakamura et al.(2001)]{Netal01} Nakamura, T., Mazzali, P.~A., 
 Nomoto, K., \& Iwamoto, K.\ 2001, \apj, 550, 991
\bibitem[Paczy\'nski(1998)]{Pa98} Paczy\'nski, B.\ 1998, \apjl, 494, 45
\bibitem[Raymond(1979)]{raymond79} Raymond, J.\ C.\ 1979, \apjs, 39, 1
\bibitem[Salpeter(1955)]{Sa55} Salpeter, E.~E.\ 1955, \apj, 121, 161
\bibitem[Schaerer \& de Koter(1997)]{Sc97} Schaerer, D., \& de Koter, 
 A.\ 1997, \aap, 322, 598
\bibitem[Skillman(1985)]{Sk85} Skillman, E.\ D.\ 1985, \apj, 290, 449
\bibitem[Snowden et al.(2001)]{Sn01} Snowden, S.\ L., Mukai, K., 
 Pence, W., \& Kuntz, K.\ D.\ 2001, \aj, 121, 3001
\bibitem[Stetson et al.(1998)]{St98} Stetson, P.\ B., et al.\ 1998, 
 \apj, 508, 491
\bibitem[Turatto et al.(2000)]{Tetal00} Turatto, M.~et al.\ 2000, \apjl, 
  534, L57
\bibitem[Wang(1999)]{Wa99} Wang, Q.\ D.\ 1999, \apjl, 517, L27  
\bibitem[Wang \& Helfand(1991)]{WF91} Wang, Q., \& Helfand, D.\ J.\
 1991, \apj, 370, 541
\bibitem[Wang, Immler, \& Pietsch(1999)]{WIP99} Wang, Q.\ D., Immler, 
 S., \& Pietsch, W.\ 1999, \apj, 523, 121 
\bibitem[Williams \& Chu(1995)]{WC95} Williams, R.\ M., \& Chu, Y.-H.\
 1995, \apj, 439, 132
\bibitem[Yang et al.(1996)]{YCST96} Yang, H., Chu, Y.-H., Skillman, 
 E.~D., \& Terlevich, R.\ 1996, \aj, 112, 146
\end{thebibliography}
\end{document}